\documentstyle[aps,floats,epsf,twocolumn]{revtex}
 
\begin{document} \draft

\title{Critical exponents at the superconductor-insulator
transition in dirty-boson systems}

\author{Igor F. Herbut}

\address{Department of Physics, Simon Fraser University, 
Burnaby, British Columbia, \\
 Canada V5A 1S6\\}\maketitle

\begin{abstract}
 I obtain the inverse
of the correlation length exponent at the superfluid-Bose glass  quantum
critical point as a series in small parameter
$\sqrt{d-1}$, with $d$ being the dimensionality of the system, and 
compute the first two terms. For $d=2$ 
I find $\nu_{s}=0.81$ and $\nu_c =1.03$,  for 
short-range and  Coulomb interactions between bosons,
respectively. When combined 
with the exact values of the dynamical critical exponents,
these results are in quantitative agreement with the experiments on
onset of superfluidity in $^4 He$ in porous glasses,
and on superconductor-insulator transition in homogeneous metallic films.

\end{abstract}

\vspace{10pt}

    The phenomenon of superconductor-insulator (SI) transition occurs in
    plethora of low-dimensional electronic systems,
    examples ranging from Josephson junction
    arrays \cite{zant} and homogeneous thin films \cite{liu}, \cite{yazdani}
    to high-temperature superconductors \cite{doniach}, and
    is believed to represent a prototypical
    quantum (zero-temperature) phase transition. At low temperatures, as
    some controlling parameter like thickness of a film 
    is varied, the resistivity
    changes from a sharply decreasing to a continuously
    increasing function of
    temperature \cite{goldman}. The good collapse of the resistivity
    data under scaling and near-universality
    of the critical value of the conductivity indicate a quantum
    ($T=0$) critical point that separates  
    two many-body ground states with different symmetries. A natural
    question arises: what is the mechanism of destruction
    of the superfluid ground state in a disordered  
    system? One possibly universal answer is provided by
    the so-called dirty-boson theory, which postulates that it is the loss
    of the superconducting phase coherence due to localization of
    Cooper pairs which is ultimately responsible for the
    SI transition \cite{ma}, \cite{rama}, \cite{fisher}.
    Since on the scale of the diverging phase-coherence
    correlation length Cooper pairs will appear as point
    particles, the SI transition would, under
    this hypothesis, in general fall  into the same universality class
    as the onset of superfluidity in 
    $^4 He$ in disordered media \cite{crowell}, corrected for the
    long-rangeness of the Coulomb interaction. In
    principle, a way to assess the validity of this 
    idea is to compare the measured critical exponents with the
    calculations for the dirty-boson Hamiltonian. A  
    strongly-coupled nature of the dirty-boson critical point, however, 
    poses a fundamental obstacle to this procedure,  and 
    makes  any but most qualitative
    understanding of the superfluid-Bose glass transition very difficult.
    The absence of a useful non-interacting starting point for 
    disordered bosons forces one to rely
    on uncontrollable approximation schemes 
    or turn to  numerical calculations \cite{herbut1}.
    This seems to be a common problem for the
    theories of interacting disordered
    low-dimensional quantum systems, apparent also in the 
    fermionic systems of this type  \cite{belitz}. In fact,
    the dirty-boson Hamiltonian may be the simplest
    quantum problem that irreducibly contains the
    physics of interactions and disorder \cite{fleishman}, and as such has
    received a lot of attention through the years \cite{herbut1}.

  Recently, a new approach to the dirty-boson criticality has been
  suggested \cite{herbut}, according to
  which the strongly coupled superfluid-Bose glass critical point in two
  dimensions ($d=2$) could be  understood as smoothly evolving from the
  zero-disorder critical point in $d=1$. The 
  idea is to note that, by preventing the  clean
  superfluid ground state in $d=1$ to exhibit a true long-range
  order, the Mermin-Wagner theorem \cite{mermin}
  forces the SI fixed point in
  $d=1$ to lie precisely at {\it zero} disorder \cite{giamarchi}. In
  $d=1+\epsilon$ a true long-range order becomes possible 
  and the superfluid thus becomes more resilient to disorder,
  which causes  
  the SI fixed point to shift to a finite, but small, value of disorder,
  controlled by the parameter
  $\epsilon$ \cite{herbut}. Although the dirty-boson
  transition probably lacks the upper critical dimension \cite{fisher1},
  it has the lower critical dimension $d_l =1$, and this in principle
  allows one to compute the universal quantities
  at the transition perturbatively in small parameter
  $\epsilon=d-1$. Within this formalism, a  particular symmetry of the
  low-energy action present in $d=1$ guarantees that the dynamical
  critical exponent is $z=d$
  ($z=1$ for Coulomb interaction) exactly \cite{herbut},
  in agreement with the expectation based on the
  compressible nature of the Bose-glass  \cite{fisher1}. The second,
  correlation length exponent $\nu$, however, turns out to be 
  a perturbation series in $\sqrt{\epsilon}$.  On the experimental side,
  a directly measurable quantity is typically 
  the product of the two exponents $z\nu$ \cite{goldman}, \cite{crowell}, and 
  a meaningful comparison with experiment requires knowledge
  of the exponent
  $\nu$ to some accuracy. In this Letter I present  a field-theoretic
  method for higher-order calculations within the $\epsilon$-expansion
  for the superfluid-Bose glass transition, and use it to compute 
  the correlation length exponent to {\it two} lowest orders in
  $\sqrt{\epsilon}$. The result for both short-range and Coulomb
  interaction between bosons (see Table I) leads to values of $\nu$ 
  in $d=2$ in a very  good agreement with the experiments on the 
  onset of superfluidity in $^4 He$ in aerogel
  \cite{crowell} and on the SI transition
  in thin metallic films \cite{goldman}, as well as with the Monte Carlo
 calculations on the dirty-boson Hamiltonian
 \cite{wallin}. My calculation supports the idea that the
  SI transition in disordered electronic systems 
  falls into the dirty-boson universality class, and establishes a way 
  for a quantitative understanding of the SI criticality, as presently
  exists for the thermal critical phenomena \cite{binney}. The 
  effort involved in higher-order calculations
  of the correlation length exponent and of the universal
  critical conductivity is discussed.

\begin{table}

\begin{center}

\begin{tabular}{|c|c|c|c|} 
&  Experiment &  Monte Carlo \cite{wallin} & $\epsilon$-expansion \\ \hline \hline
$\nu_{s}$ & $0.80\pm 0.04$ \cite{crowell} & $0.90\pm 0.10$ & $0.81$ \\ \hline
$z_{s}$ & $2$ & $2.0 \pm 0.1 $ & $2$ \\ \hline
$\nu_c$ & $1.2\pm 0.2$ \cite{nina} & $0.9\pm 0.15 $ & $1.03$ \\ \hline
$z_c$  & $1.0\pm 0.1$ \cite{yazdani} & $1.0$ & $1$ 

\end{tabular}
\end{center}

\caption[]{Comparison between experiment, Monte Carlo, and
second-order $\epsilon$-expansion results for the critical exponents, for
short-range (s) and Coulomb (c) universality classes.}

\label{table1}
\end{table}

To be specific, consider the effective low-energy $T=0$ action for the
disordered superfluid in $d=1$ \cite{haldane}, \cite{fisher1}:
\begin{eqnarray}
S= \frac{K}{\pi} \sum_{i=1}^{N} \int_{-\infty}^{\infty} dx d\tau \{
c^2 [\partial_x \theta_i (x,\tau)]^2 + [\partial_\tau \theta_i (x,\tau)]^2 \}
 \\ \nonumber
-D \sum_{i,j=1}^N \int_{-\infty}^{\infty}
dx d\tau d\tau' {\rm cos} 2[\theta_i (x,\tau) - \theta_j (x,\tau ')].
\end{eqnarray}
The coupling $K$ is inversely proportional to
the superfluid density at some microscopic cutoff length $\Lambda^{-1}$,
$c$ is the velocity of
low-energy phonons, and $D$ is proportional to
the width of the Gaussian 
random potential. The interaction between particles in (1) is taken  to be
short-ranged, and the standard limit on the number of replicas
$N\rightarrow 0$ is 
assumed. The field $\theta(x,\tau)$ is the {\it dual} phase
\cite{haldane}, \cite{herbut1}, and the above
theory describes the destruction of the second sound mode
in the superfluid due to unbinding  of topological defects
(phase slips) 
at the point of transition in $d=1$. Its role is similar to that of 
the sine-Gordon theory for the Kosterlitz-Thouless
transition in the 2D XY model \cite{amit}. To determine the
macroscopic state of the system, one is in principle 
interested in fate of the couplings $K$, $c$, and $D$ as the
cutoff in the theory is lowered. 
Under a change $\Lambda \rightarrow
\Lambda/s$ the combinations of the
coupling constants $u=3-\eta^{-1}$, (where $\eta=Kc$),
$\kappa=1/Kc^2$, 
and $W\sim D$ (to be precisely defined shortly),
in $d=1+\epsilon$ dimensions are expected to 
renormalize according to the $\beta$-functions \cite{herbut}:
\begin{equation}
\dot{u} = \epsilon (u -3) + W + a u W + O(W^2),
\end{equation}
\begin{equation}
\dot{W}= u W + b W^2 +O(u W^2), 
\end{equation}
\begin{equation}
\dot{\kappa} = (d-z) \kappa,
\end{equation}
 where $\dot{x} = dx/d\ln(s)$, and $a$ and $b$ are numerical
 coefficients.  The $d$-dependent terms in the recursion relations
 (2)-(4) can be understood as deriving from the  
 scaling of the superfluid density, $\rho_{sf}\sim K^{-1} \sim
 \xi^{2-z-d}$, and the compressibility, $\kappa\sim \xi^{z-d}$
 \cite{herbut}, \cite{fisher1}, where $\xi$ is the diverging correlation
 length near the critical point, and $z$ the dynamical
 exponent. Adopting the  logic of the
 minimal subtraction scheme \cite{binney},
 the disorder-dependent terms in the recursion relations 
 should be computed precisely in $d=1$, where one has the dual
 representation (1) of the low-energy theory on disposal.
 The symmetry of the interaction term in (1)
 under a transformation $\theta_i (x,\tau) \rightarrow \theta_i (x,\tau)
 + f(x)$ for arbitrary $f(x)$ implies then that 
 there could be no disorder-dependent terms in Eq. (4) \cite{herbut}, so 
 $z=d$ at the fixed point. The correlation length exponent $\nu$
 follows from the linearization
 of the first two equations near the critical
 point at $W^* \sim u^* \sim \epsilon$.
 It is then straightforward to check that 
 to the second order in $\epsilon^{1/2}$  
 \begin{equation}
 \nu_{s} ^{-1} = 3^{\frac{1}{2}} \epsilon^{\frac{1}{2}}
 + \frac{1+3(a+b)}{2} \epsilon + O(\epsilon^{\frac{3}{2}}), 
 \end{equation}
 for short-range interactions. The $O(\epsilon^{3/2})$ term follows
 from the higher-order terms in Eqs. (2)-(3).

  While the outlined procedure is conceptually
  simple, its implementation is made difficult by the fact that the
  action (1) has a compact form only in real space, and the
  interaction term contains
  all powers of the dual field. A similar obstacle  appears 
  in the calculation of the Kosterlitz recursion
  relations beyond the lowest order in fugacity \cite{amit}.
  Here I will introduce
  a field-theoretic approach to the problem, which also enables one 
  to avoid the usual pitfalls of the momentum-shell renormalization
  group when applied beyond the lowest order.
  The gist of the method
  is to recognize that in $d=1$, right at $u=0$ the 
  coupling constant $W$ becomes dimensionless, and the theory (1)
  appears to be just renormalizable. One then expects that the
  logarithmic divergences at the renormalizable point  $d=1$ and
  $u=0$ will at small
  finite $u$ show  as poles when $u\rightarrow 0$. Since the coupling
  $W$ acquires a finite
  scaling dimension for $u\neq 0$, the coefficients in the
  $\beta$-functions are expected to
  stay finite as $u\rightarrow 0$. This is analogous
  to the standard procedure of dimensional regularization, commonly used
  to study thermal critical phenomena near the upper critical dimension
  \cite{binney}, except that here the coupling constant $u$ plays the role
  of dimension, while the real physical dimension is at first 
  fixed at $d=1$. When finally $d \rightarrow 1+\epsilon$, the
   $\beta$-functions are deformed into Eqs. (2)-(4). 

  With this strategy in  mind, consider the self-energy defined by  the
  propagator of the dual phase in $d=1$ as $G^{-1} (k, \omega)
  = (2K/\pi) (\omega^2 + c^2 k^2 ) +\Sigma (\omega)$. It will prove
   useful to
  separate the first and the second order contributions to $\Sigma$
  by writing it as
  $\Sigma(\omega) = \Sigma_1 (\omega) + \Sigma_2 (\omega) + O(D^3)$,
  where $\Sigma_n \sim D^n$. Simple calculation then gives
  \begin{equation}
  \Sigma_1 (\omega)
  = 8 D\int_{-\infty}^{\infty}d\tau (1-{\rm e}^{i\omega\tau})
  e^{-f(\tau)},
  \end{equation}
  where the two-point correlation function
  $f(\tau) = 4\langle \theta(0,0) ( \theta(0,0) -\theta(0,\tau))\rangle$
  is given by the integral
  \begin{equation}
 f(\tau) = (3-u) \int_0 ^{c\Lambda|\tau|} \frac{dx}{x}(1-e^{-x}).
 \end{equation}
 When $u\rightarrow 0$, it readily follows that at small frequencies 
 \begin{equation}
 \Sigma_1 (\omega) = \omega^2 \frac{2W}{\pi c } (\frac{1}{u}+ O(1)),
 \end{equation}
 where I introduced the
 frequency-dependent, dimensionless coupling 
 $W= ( 4 \pi D/( c^2 \Lambda^3 ) )
(c\Lambda/\omega)^u $. After a tedious but straightforward algebra
one similarly finds
\begin{equation}
\Sigma_2 (\omega) = \frac{\pi}{2K} (\frac{ \Sigma_1(\omega)}{\omega} )^2 +
I(\omega), 
\end{equation}
where
\begin{eqnarray}
I(\omega)= 8 D^2 \int_{\infty}^{\infty} dy d\tau d\tau' dv
(1-e^{i\omega\tau})
{\rm e}^{-f(\tau) -f(\tau')} \\ \nonumber
[ F(y,v,\tau,\tau')(1+{\rm e}^{-i\omega v}
\frac{1}{2}(1-{\rm e}^{-i\omega
\tau'}) -1],  
\end{eqnarray}
and $F$ denotes a four-point correlation function:
\begin{equation}
F(y,v,\tau,\tau')={\rm e} ^{-4 \langle (\theta(0,0) - \theta(0,\tau) )
(\theta(y,v)-\theta(y,v+\tau'))\rangle }. 
\end{equation}
 When $\omega\rightarrow 0$, after rescaling the imaginary times
  and the length in the integral  as $\omega\tau \rightarrow \tau $ and
 $\omega y/c \rightarrow y$, the leading divergence
 in $I(\omega)$ as $u\rightarrow 0$ comes from the integration over
 small values of $\tau$ and $\tau '$. To obtain the leading
 divergence in $I(\omega)$ it therefore suffices 
 to expand $F$ to the lowest  order
 in $\tau$ and $\tau '$, to find that at small frequencies 
 \begin{equation}
 I(\omega)= -\omega^2 \frac{6}{\pi c } W^2 (\frac{1}{u^2} + O(\frac{1}{u})).
 \end{equation}
 The last equation is the central result of this work.
 Collecting all the terms back into the self-energy one
 recognizes the renormalized  coupling
 $\eta_r$ as the coefficient of $\omega^2$-term in the propagator.
 In general,
 \begin{equation}
 \eta_r = \eta + \frac{W}{u} + x \frac{W^2}{u^2} +O(\frac{W^2}{u}),  
 \end{equation}
 where the terms finite when $u\rightarrow 0$ have been discarded, and
 $x$ is a number determined from Eq. (12).
 After judiciously defining a renormalized disorder in (13)
 as $W_r= W + 2 x W^2 /u$,
 and rescaling it to bring the coefficient of $O(W)$-term
 in the Eq. (2) to unity as  
 $9 W_r \rightarrow W_r $, a differentiation with respect
 to $\ln(c\Lambda/\omega) $ leads to the
 Eqs. (2)-(3) for the renormalized couplings $u_r$ and $W_r$,
 with the coefficients $a=-2/3$ and $b=2 x/9$, with  
 $b=0$. The subleading $\sim W^2 /u$
 term in the Eq. (13) determines the next, $O(W^2)$, term in (2), and 
 the next-order correction to $\nu_{s}^{-1}$.

      A remarkable feature of the perturbation series for $\nu_{s}$ is its
 independence of the renormalization procedure, i. e. of the
 non-universal finite parts of the self-energy
 which have been dropped in the last equation. To see this 
 to the order of my calculation  consider the  most general
 redefinition of the coupling constants to the order $W^2$ \cite{amit}:
 \begin{equation}
 u_r ' = u_r +\alpha W_r +\beta u_r W_r +\gamma W_r ^2, 
 \end{equation}
 \begin{equation}
 W_r ' = W_r +\delta u_r W_r + \sigma W_r ^2, 
 \end{equation}
 where the coefficients $\{ \alpha, ... \sigma \} $ are finite, and
 dependable on the finite parts of the self-energy. 
 It is easy to check that the recursion relations for
 the new couplings have the same form as the Eqs. (2)-(3), but
 with the coefficients $a' = a+\alpha - \delta $ and $b' = b -\alpha
 +\delta $. Interestingly, while the coefficients $a$ and $b$ by themselves
 are non-universal, the critical exponent depends only on their sum,    
 which is perfectly universal, i. e. scheme independent. An 
 invariant similar to $a+b$ appears also in the $\beta$-functions for the 
 sine-Gordon model \cite{amit}, where it  
 determines the first correction to Kosterlitz-Thouless scaling.  

  I expect the presented $\epsilon$-expansion
  to lead to  divergent series; the point
   $D=0$ in the action (1) is non-analytic, since for $D<0$
   the Gaussian probability distribution for the random potential becomes
   unbounded. Nevertheless, the hope is that the series
   in Eq. (5) for example, will be asymptotic, and that 
   the few lowest terms may already lead to useful results. Indeed, 
   estimating $\nu_{s}$ for $d=2$ from the simple sum of the
    first two terms gives $\nu_{s} =0.81$, within bounds found
 in the Monte Carlo calculations \cite{wallin}.
 The experimental data of Crowell et al. on the onset of superfluidity and
 on the specific heat of $^4 He$ in aerogel at low temperatures \cite{crowell}
 on its face value are consistent with the effective dimensionality
 of $d=2$. Under this assumption the product of the two
 exponents in their experiment is $z\nu = 1.60 \pm 0.08$.
 Assuming further that $z_s =2$ at the transition in $d=2$
 gives $\nu=0.80 \pm 0.04$. Although the uncertainty cited
 here should be taken
 with some reservation, and the accuracy of the measurement 
 falls short of the standards in thermal critical phenomena, the result 
 appears to be  in excellent
 agreement with my calculation (see Table I). It is worth noting 
 that the inequality \cite{chayes} $\nu \geq 2/d$
 seems to be violated both by the experiment and by my estimate.
 It has been argued recently \cite{pazmandi} that the above
 inequality is an artifact of the particular averaging procedure, and that
 the true exponent is in fact not bound from below.
 It would be interesting to see if the higher-order corrections 
 eventually  push the value of $\nu_{s}$ above unity in $d=2$,
 or indeed $\nu_{s} < 1$ as the experiment and the present
 calculation suggest.

  To make a comparison with the experiments on SI
  transition in homogeneous thin films with thickness as the tuning
  parameter \cite{liu} it is
  necessary to take into account the long-range Coulomb interaction
  between the Cooper pairs. As explained in detail elsewhere
  \cite{herbut}, \cite{herbut2} within the present scheme 
  this may be simply accomplished by defining the Coulomb  interaction
  as $V_c (\vec{r}) = e^2 \int d^d \vec{q} \exp(i\vec{q}\cdot\vec{r})
  /q^{d-1} $, so to coincide with the $\delta$-function repulsion
  in $d=1$. The only change in the calculation then is that
  $z_c =1$, and that $\epsilon\rightarrow
  \epsilon/2$ in the Eq. (2), while the disorder-dependent terms
  in the recursion relations, which follow from $d=1$, remain the same.
  The first two terms in the series
  then give $\nu_c = 1.03$, in accord with the 
  Monte Carlo results \cite{wallin} (see Table I).
  Experiment finds $z_c =1.0 \pm 0.1 $, by
  suppressing the transition temperature
  with the magnetic field or 
  by scaling of resistance with the electric field
  \cite{yazdani}.  Collapsing the resistivity data 
  \cite{nina} then gives the experimental value of
  $\nu_c = 1.2 \pm 0.2$, again in a very good agreement with my result.

  As mentioned earlier, the next term in the series (5) requires
  only the computation of  the subleading, 
  $O(W^2 /u)$ term in Eq. (13).
  Here, however, it appears that it is no more enough to know the
  correlator $F$ (after rescaling the lengths with $\omega$) only at
  small rescaled $\tau$ and $\tau '$, as it was for the
  leading divergence in Eq. (12). In light of the likely asymptotic
  nature of the expansion, this is left for future work.

  Another universal quantity of interest is the
  critical conductivity in $d=2$ \cite{fisher},
  which, aside from the universal unit
  $e_* ^2 /h$, for bosons of charge $e_*$
  can be obtained as a Laurent series
  in $\epsilon$ \cite{herbut2}. The lowest order term was obtained
  in \cite{herbut2}, and for $d=2$ the result $\sigma_c = 0.69 e_* ^2 /h$
  agrees with the low-temperature experiment on Bismuth films
  \cite{nina} quite well. It would
  nevertheless be useful to compute the next-order correction,
  and see if it pushes the result towards the self-dual value of
  $e_* ^2 /h$, to which a large number of experimental results seems to
  converge. Calculating the next-order term  in the critical conductivity 
  would in principle proceed along the same lines as here,
  except that one needs  
  to perform the analytic continuation to real frequencies to obtain the 
  real-time dc response at low temperature. This problem was solved
  in \cite{herbut2} to the lowest order, but applying the same trick
  in the next
  order term seems not straightforward.  A preliminary analysis also 
  suggests that the second-order contribution
  to the universal conductivity requires 
  a computation of both leading and subleading divergences in the
  second-order contribution to the self-energy.

   In conclusion, it is shown that the expansion around the
   lower critical dimension $d_{l} =1$
   for the superfluid-Bose glass critical point allows a 
   field-theoretic formulation that facilitates a systematic
   higher-order calculation of the critical exponents. The
   computation to two lowest orders yields results for the correlation length
   critical exponents in $d=2$ in respectable agreement with the experiment
   and Monte Carlo calculations, both for the short-range and the Coulomb
   interaction between particles. The results suggest that 
   the superconductor-insulator transition in homogeneous
   thin films is in the universality class of disordered bosons. 

   This work has been supported by NSERC of Canada.

\end{document}